Rev: 1/10/2023

Pragmatic Estimation of Sample Size for Number of Interviews for PRO development in the 2009 FDA PRO guidance

Affiliation

Chris Barker, Ph.D.

Adjunct Associate Professor of Biostatistics

University of Illinois Chicago - School of Public Health

Chicago, Illinois

And

Chris Barker, Ph.D.

CEO

Chris Barker Statistical Planning and Analysis Services Inc.

www.barkerstats.com

Abbreviations

CRC – Capture-Recapture

KM – Kaplan Meier

PRO – Patient Reported Outcome

FDA-Food and Drug Administration

# Abstract

Patient Reported Outcomes developed according to the 2009 FDA PRO guidance require an initial step of structured patient interviews or focus groups. The guidance does not provide sample size suggestions or methodology. This paper proposes statistical methodology and sample size guidance that address this gap in the FDA PRO guidance. This paper also appears to be the first to provide a definition of Type I error in interviews for a PRO methods for assessing sample size and a new definition of saturation based on a probability distribution to confirm whether enough interviews have been prepared. Type I error is declaring saturation when it has not been achieved during the interviews. Two worked examples applied to actual interview data and published data, are presented. Guidelines are proposed for the estimation of sample size useful for PRO experts conducting interviews for a PRO. These methods are applied in the setting of qualitative research. Future research for other methods of interviewing and determining saturation is required.

# Background

This paper and methodology arose because of a project progress teleconference with the CEO and CMO for a small biotech, me, a PRO developer vendor and additional clinical operations staff. The vendor was responsible for activities including interviewing patients, preparing and validating a PRO instrument for use as a primary endpoint in a Phase III randomized trial according to the FDA PRO guidance (FDA, 2009). Due to confidentiality agreements, the names of the biotech, vendor, drug, and indication are withheld.

The CEO asked the vendor how many interviews would be required to develop the instrument. The vendor was processing the interviews using Qualitative Research and stated that interviews would be complete "once saturation was achieved". The CEO requested I explain the definition of "saturation". The vendor defined 'saturation' as the first interview that elicited no new concepts. I asked a follow-up question "Is that the first occurrence of saturation or do you conduct several, perhaps 2 or more additional consecutive interviews with no new concepts?". The vendor replied "no, only one interview with zero new codes". I also asked if a single interview could be statistical noise and possibly a type I error (declaring saturation when it had not occurred). The vendor was unable to answer that question about Type I error.

Excerpting from the PRO guidance (page 13), "We cannot provide recommendations for the number or size of the individual patient interviews or focus groups for establishing content validity. The sample size depends on the completeness of the information obtained from analysis of the transcripts. Generally, the number of patients is not as critical as interview quality and patient diversity included in the sample in relation to intended clinical trial population characteristics.". The methodology of this paper addresses the number of patients but does not address the interview quality or patient diversity.

Qualitative research is one of several methodologies for preparing and coding interviews used to develop PRO's. An overview of other methods is described in a paper sponsored by ISPOR, and includes, phenomenology, Grounded Theory, etc. (Patrick 2011). The implementation of the interview methods is not reviewed in this paper. This paper considers only qualitative research implements the empirical concept of 'saturation' to determine when interviews are complete. In reviewing the literature about the concept of saturation there does not appear to be a formal statistical method for determining either saturation or statistical guidance for sample size. Note the applicability of the methods of this paper to other methods of conducting interviews must be assessed on a case-by-case basis. Aside from the case-by-case assessment a common property of the other methods is the range of number of interviews with "zero new codes".

This paper does not provide a complete literature review of the concept of saturation, an ambitious goal well beyond the scope of this manuscript. The term "saturation" is frequently applied in grounded theory research and its application beyond that theory is debated (Oreilly 2013). Theoretical saturation means that researchers reach a point in their analysis of data that sampling more data will not lead to more information related to their research questions (Seale,1999). I mention a small number of arbitrarily selected papers have reviewed interview methodology, definition of saturation and saturation and the number of interviews (Francis, 2010, Fusch 2015, Mason 2010). The term "saturation" is defined as zero new concepts (or themes) elicited at an interview. Alternate definitions appearing in the literature are: first occurrence of saturation, three consecutive interviews with saturation, expert Judgement "additional interviews would be counter-productive", minimum 10 interviews and 3 with zero new codes ("10+3"), (Francis, 2010).  Several deterministic recommendations for number of interviews (sample size) appear in the literature.

Table 1

| | |
|---|---|
| Ethnography/ethnoscience | "30 -60 interviews" |
| Grounded Theory | "30-50 interviews" |
| Phenomenology | "5-25 interviews" |
| All Qualitative research | "at least 15 interviews" |
| Funded ($) research 'time limited' interviews ranged | from 1 – 95 |

Tesch (1990) enumerates 23 "qualitative research types" for which the methods of this paper may also be applicable.

# Methods

## Assumptions for application of the methodology of Kaplan-Meier.

Throughout the discussion I use the term "sample size" as synonymous with number of interviews. I assume that there is one subject per interview (subjects do not give more than one interview). The Kaplan Meier methodology adopted here for qualitative research  is descriptive (no "p-values"). One major property of the Kaplan Meier as defined here, the KM probability of not being saturated is unchanged or declines with each successive interview. The KM probability estimates can decline and exactly equal zero.  A further characteristic is that at and after saturation, additional interviews are redundant and do not provide new codes.

This paper assumes there is a fixed but unknown number of "codes".  One must assume that each interview is conducted in a repeatable reproducible manner. Each interview results in none, one or more new codes vs. preceding interview(s). The Kaplan-Meier assumes that each outcome or event "new codes" or "no new codes", in this example each interview, is statistically independent. This is unlikely to be true for interviews.  The fixed unknown total number of codes induces a correlation among pairs of interviews. Modelling this correlation is outside the scope of this paper. The Kaplan-Meier estimate of probability of saturation is unbiased. Due to the induced correlations, the estimates of variability may be  biased, and may tend to be too small relative to the independence case. Alternative estimates of variability such as by using a Bootstrap might be applicable and are not further examined in this manuscript.

## Adaptation of KM methodology to other interview methodologies.
The KM methodology must be adapted for each interview methodology.

# Example data sets and availability

These datasets below are available on request. Informed consent was obtained from the subjects providing data for peach of the datasets. The Guest data used below was collected as part of a larger study (Guest 2006). Informed consent for the subjects was obtained in the larger study (Bernard, Personal communication, 2020). The patients for the second dataset gave informed consent (Revicki, Personal communication, 2017). Table 1 presents a "flat file" structure for organizing the data. I recommend that the interview datasets are included in the regulatory submission as part of standardized ADaM and SDTM datasets. The FDA currently requires clinical trial data submitted in this standardize format (Wood, 2008).

## Data set I – anonymized interview data

I obtained an anonymized dataset from an expert in PRO's (Kleinman, 2012) from a set of twenty-one interviews processed using qualitative research. The anonymization replaced the code description with a letter (A, B...). Interviews were in chronological order and a 1 represents the code observed at the interview 0 otherwise. The interviews resulted in 20 separate codes.

The dataset may be represented in a tabular format in table 2. Aligning with conventional terminology for capture-recapture discussed below the "codes" are termed "marked" (M) when first elicited, and when a code is elicited a second or later interview, it is labelled a "recapture" (R). For each, code the 0's represent code not elicited in interview n and 1=code elicited in interview j. Table 3 summarizes the data.

Table 2- Organization for Interview Dataset

| Interview ID | Interview Sequence # | Anonymized Codes (M=Elicited, R= code Elicited again) | | | | |
|---|---|---|---|---|---|---|
| | | Code A, | Code B, | Code C, | ... | Code K, |
| | | (M,R) | (M,R) | (M,R) | (M,R) | (M,R) |
| aaaa | 1 | (1, 0) | (0 ,0) | (0 ,0) | | 0 ,0) |
| bbbb | 2 | (0 ,0) | (0 ,0) | (1,0) | | (1,1) |
| cccc | 3 | (1 ,1) | (1,0) | (0 ,0) | | (0 ,0) |
| dddd | 4 | (1 ,1) | (1,1) | (0 ,0) | | (0 ,0) |
| eeee | 5 | (1 ,1) | (1 ,1) | (0 ,0) | | (1 ,1) |
| ... | ... | | | | | |

The data set has the following raw and derived variables,

Let $j$ = interview chronological sequence number, j=1,2,3.... $J$ where $J$ is the total number of interviews

Note that total number of interviews, J is unknown at the start of the interviews.

Code -k-, (E)licited at interview j, $E_{jk}$: 1=yes, 0=no

Where k=1,2, 3... $K$, elicited from interview, j=1,2, 3... $J$

Note that the total number of codes $\mathbb{K}$ is unknown at the start the interviews.

Let $R_{j,k}$ indicate $E_{j,k}$ elicited again "recaptured" at interview, j ,(1=yes, 0=no k=1,2, 3… $\mathbb{K}$ ),

total codes elicited at interview $\mathbf{j}$, $N_j = \sum (E_{j,k})$ and summation over k cumulative elicitations of code k, $M_k = \sum E_{j,k}$ summation over j.

| Table 3 interview descriptive statistics | | | | | | | | | |
|---|---|---|---|---|---|---|---|---|---|
| | **N marked codes** | **Mean marked per interview** | **Median marked per interview** | **STD marked** | **N recap** | **Mean recapture** | **Median recapture** | **STD recapture** | |
| | 47 | 4.19 | 4.00 | 2.34 | 47 | 3.81 | 3.00 | 2.44 | |
| | | | | | | | | | |

| Table 4 interview descriptive statistics recaptured | | |
|---|---|---|
| | **Recapture** | **N** |
| | 0 | 2 |
| | 1 | 6 |
| | 2 | 8 |
| | 3 | 11 |
| | 4 | 3 |
| | 5 | 4 |
| | 6 | 5 |
| | 7 | 5 |
| | 8 | 1 |
| | 9 | 1 |
| | 10 | 1 |

## Data set II – using a Published figure from Guest et al

I conducted a search in google and located a paper by Guest et al with codes elicited from interviews then used to develop an instrument about HIV in subjects in Africa (Guest . 2006).

I contacted Dr. Guest and the raw data for his paper are not available. Figure 1 below reproduces figure 1 of the Guest paper.  Guest summarized their interviews and reported number of codes elicited in groups of 6 consecutive interviews and for interviews from Ghana then from Nigeria.  For the estimate of probability of saturation presented below a simple ad hoc imputation of codes to interviews was implemented. When #codes > 6 (interviews) then, it was assumed every interview had at least 1 code. When #codes <6, for example 4, it was assumed that four interviews resulted in 1 code and remaining 2 randomly selected interviews had zero codes.

| Figure 1 |
|---|
| Guest, reproduced with permission |

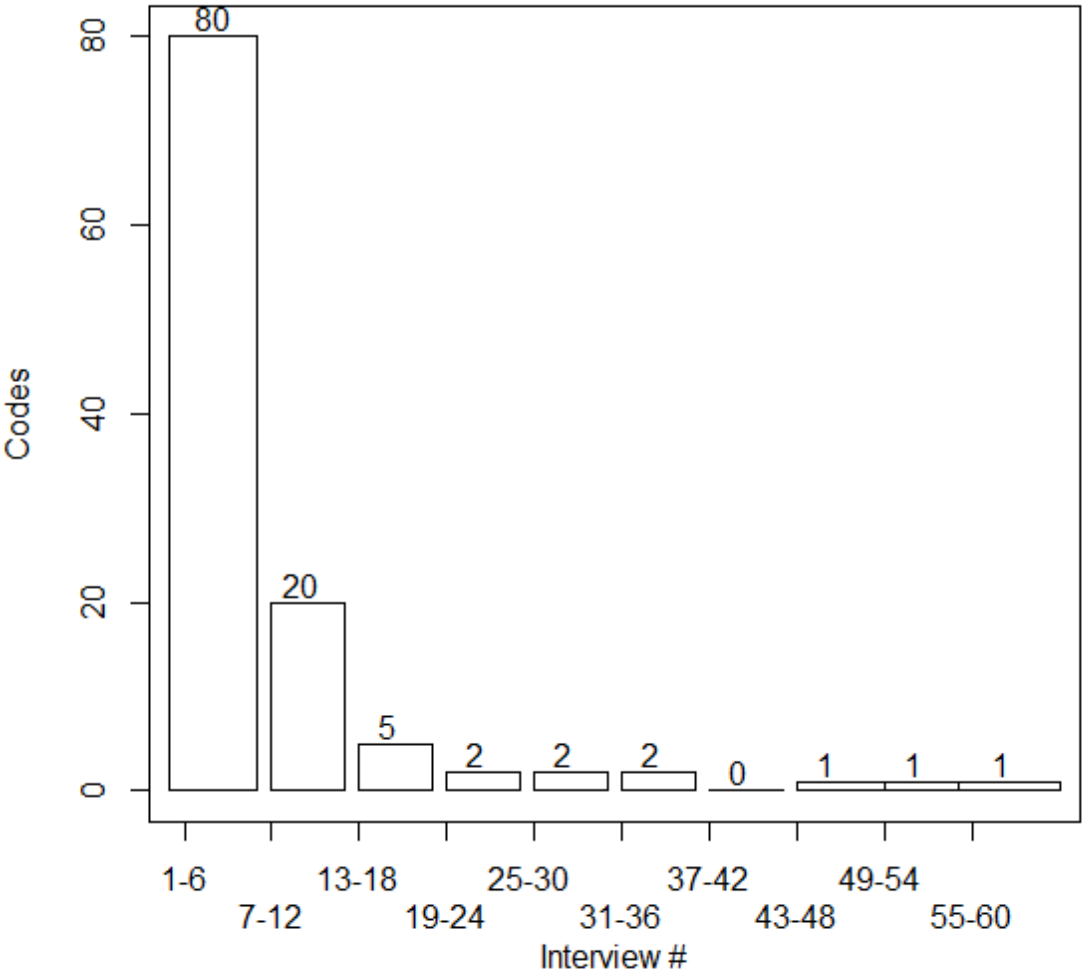

Reproduced from Guest et al with permission
Note numbers at top of bars are number of codes elicited in 6 interviews

A Type I error for PRO development may be defined as terminating interviews and incorrectly claiming saturation is achieved when more interviews are required. Consider the deterministic rule that saturation is the first occurrence of an interview with zero new codes. The data from Guest indicate that for interview 13-18, there were 5 codes in 6 interviews, therefore there was at least one interview with zero new codes. Note had interviews stopped at 13-18 would be a Type I error (saturation did not occur) , and as many as 45 more interviews were required.

Selecting a probability distribution to fit the decline in the number of "new codes" to zero can be based on a goodness of fit. By a serendipitous choice I adopted the non-parametric Kaplan Meier ("K-M") (Kaplan, Meier, 1958) . The KM does not require an assumption of a distribution. The KM is zero at an "interview" where saturation occurs. Other distributions, exponential, gamma etc. may be considered, however those distributions can equal zero. I recommend selection of a distribution that can equal zero. Consideration of adaptation of a distribution such as a truncated or triangular distribution is beyond the scope of this article.

I consider a simple application of the non-parametric Kaplan Meier estimate of the distribution of elicited codes. For the K-M "0 new codes" is treated as an "event" and >=1 code as "censored". That curve and the 95% confidence interval is presented in Figure 2. Extracting data from the graph uses the simplifying assumptions. For each group of 6 interviews, for example resulting in 5 codes, I use a simplifying assumption that one interview had zero codes and I randomly select one interview, the remaining interviews are imputed by one code per interview.

I fit a regression line to the upper 95% C.I. limits and extrapolate that to the x-axis for dataset II, presented in Figure 2. The extrapolated regression crosses the axis at approximately 70 interviews. The median is the statistic reported from the Kaplan-Meier. For saturation, the interest is the probability estimates. The KM estimate starts at 1 (100%) because no interviews have occurred - the probability that saturation has not occurred is 100%. The K-M probability estimate declines to 0.0%, at about 55 interviews, interpreted as probability 0.0% that saturation has not occurred.

I do not recommend nor adopt a hypothesis testing framework for interpreting saturation. The Kaplan Meier provides a confidence interval associated with the probability distribution to incorporate variability. As above, it may not be reasonable to assume individual interviews and elicited codes are not statistically independent. Future research in this area may consider use of bootstrap resampling or other methods that account for the potential non-independence.

## Results

Figure 3 present a Kaplan-Meier estimate of the probability of saturation, based on anonymized interview data provided by Revicki. The x-axis is the interview in chronological sequence and y axis KM probability. An "event" is an interview with zero new codes otherwise censored. The upper confidence interval is extrapolated to approximately 55 using a linear fit. The KM drops to 0 at approximately 48

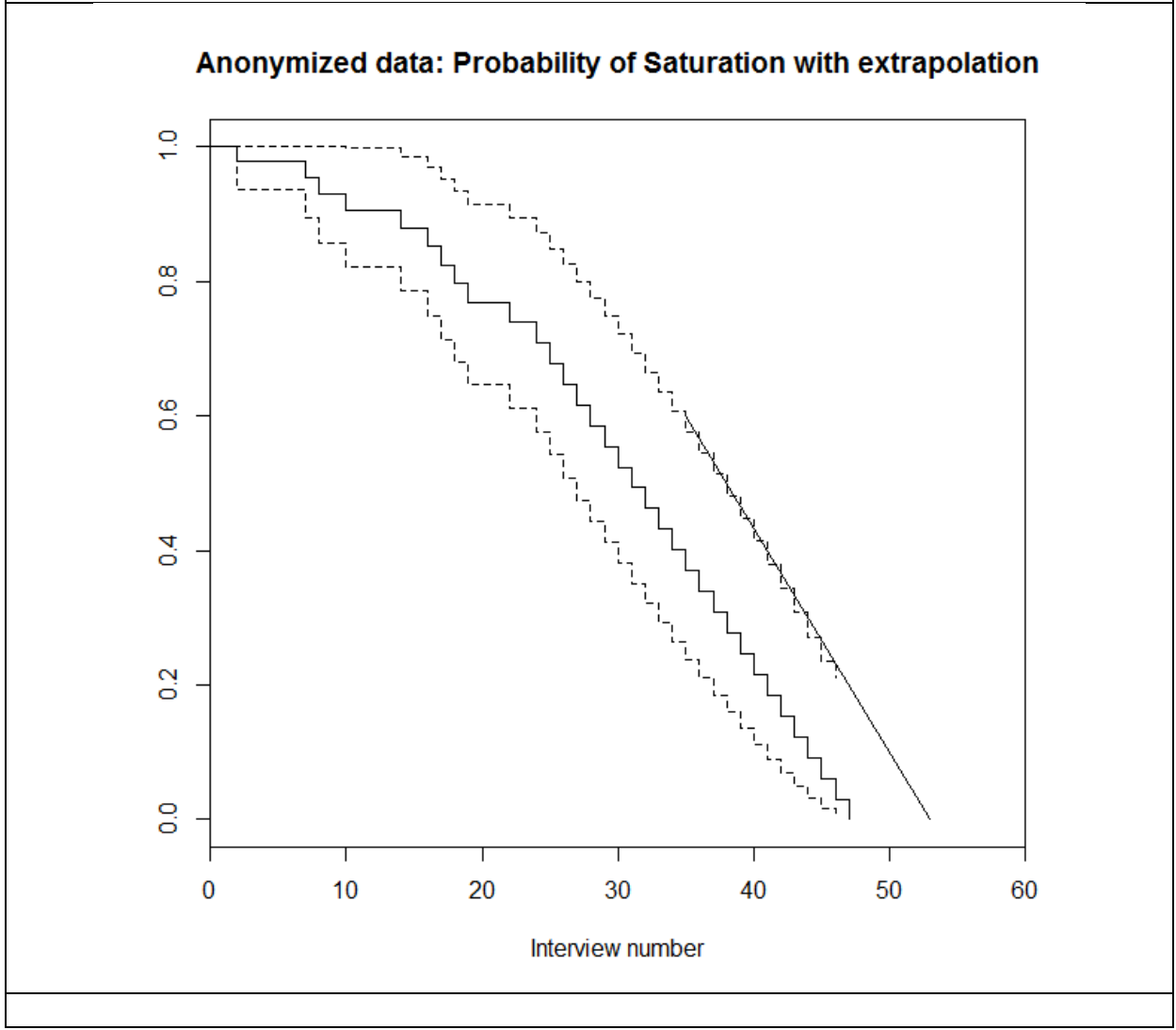

The statistical methodology presented here, fits a probability distribution to the elicited code data from Guest.  The example is simplified in part because the raw data is not available, and I extract the data from Figure 1 of Guest et all as appears in Figure 3.

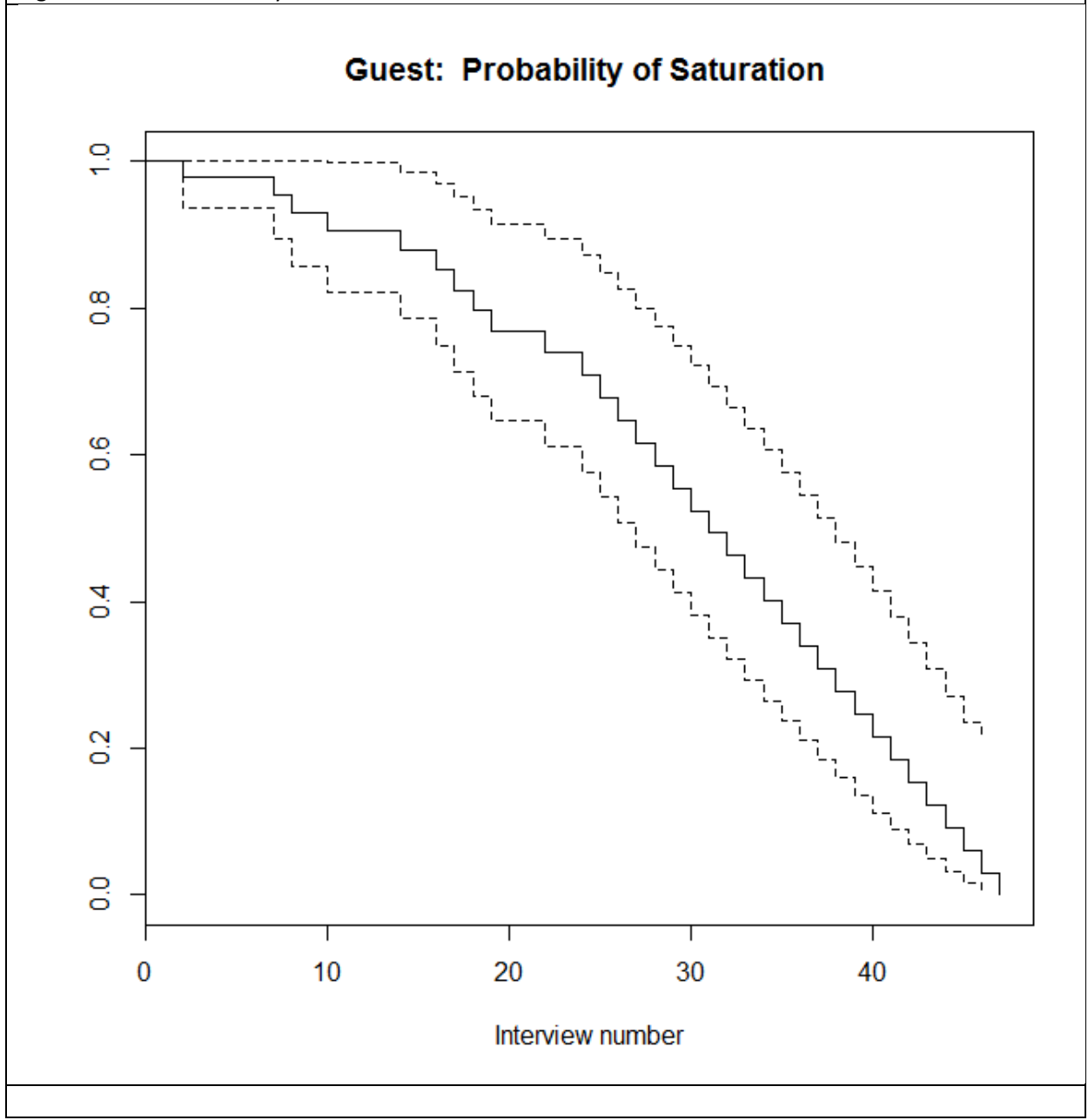

## Pragmatic Sample Size estimation

There are well established methods for estimating sample size for survival "time to event" endpoint (Wu, 2016). When sample sizes (number of interviews) are on the order of 10 to as much as 50 interviews, then a simple pragmatic "trial and error" type method may be used estimating sample size. The Kaplan Meier can be estimated in excel or in an open source language such as R (R core team, 2020). Note a reason for estimating sample size (number of interviews) could be for estimating a budget for an instrument development project. In a pharmaceutical Phase III clinical trial sample size must be completely specified prior to enrolling the first patient. For instrument development the sample size (number of interviews)  should be a  "very good guess" and allow for the possibility to be revised during the interview process.

I present several sample size planning scenarios in Table 5, to illustrate a pragmatic approach to estimating the number of interviews.  Use best available methods, such as expert judgement for an initial estimate. Then generate hypothetical sequences of interviews where each element is either 0 new codes, or 1 or more new codes.

Scenario I

As an example, suppose hypothetically the initial  expert judgement is that 10 interviews are needed. Below I denote  "1"=  interview with new codes , "0": interview with zero (0) new codes. Note ten interviews is completely hypothetical.

Scenario1: All new codes elicited in early interviews :                    (1,1,1,1,1,0,0,0,0,0)

Scenario2: All new codes elicited in the later interviews;                (0,0,0,0,0,1,1,1,1,1)

Scenario3:  Interviews eliciting new codes uniformly distributed:            (1,0,1,0,0,1,1,0,1,0)

| Table 5 | | | |
|---|---|---|---|
| Pragmatic Sample size (number of interviews) determination | | | |
| Assume base scenario of 10 interviews | | | |
| Scenario | Interview Sequence 1=new code 0= zero new codes | KM Probability of saturation (95% CI) | Additional interviews with zero new codes required for KMprob=0 |
| 1: All new codes elicited in early interviews | (1,1,1,1,1,0,0,0,0,0) | 0.5 (0.269 ,0.929) | 3 |
| 2: all new codes elicited in the later interviews | (0,0,0,0,0,1,1,1,1,1) | 0.0 ( na, na) | 0 |
| 3: Interviews eliciting new codes uniformly distributed: | (1,0,1,0,0,1,1,0,1,0) | 0.24 (0.0479 ,1) | 3 |
| | | | |

Na: not estimated

Table 5, scenario 1, presents a scenario of ten interviews, where the KM probability of (not being saturated) at the last interview is larger than zero. By a pragmatic trial and error, adding three additional interviews with zero new code gives an estimate of probability of (not being ) saturated of zero.

## Discussion and Conclusions

This paper presents a probability foundation for the definition of saturation and provides a methodology for defining Sample size in the context of the FDA PRO guidance.

PRO experts planning and conducting interviews for development of PRO's may use standard one sample statistical methods (Wu, 2016) for estimation of say, sample size required for estimation for a survival distribution.

There do not appear to be any publications that provide any summary of interview level data in the literature which could be used for sample size planning purposes.

## Probability Framework for interviews

Guest (2020) argues that there may be no probability model for saturation and cites Galvin , Fugard &Potts, and others as examples of probability-based models. Limiting the discussion to qualitative research, I propose a probability framework for saturation. The probability assumption is based on a hypergeometric distribution arising in an estimation setting called "capture recapture" (CRC) (Tilling, 2001). Capture-recapture estimates the total population (N) under the assumption there is a fixed and unknown number of "elements". Common examples of capture-recapture are in estimation of total population (Wittes, 1974) , and completeness of a disease registry, cancer. Homelessness, mental health, drug use, and in software development number of undetected software errors ( "bugs") (Chun 2006). A key assumption of capture-recapture is that codes have an equal probability of elicitation. Again this assumption may not be reasonable if successive interview questions change or evolve based on prior interview questions and answers.

I used Capture-recapture (CRC) methods to estimate the total number of "codes" for an instrument for an "indirect estimate" of sample size. The CRC would give an estimate of the number of codes. Consider an instrument such as the SF36 where there are 36 codes. An estimate at say, the second or third interview that the total number of codes is approximately 36 but at the hypothetical second interview only two 2 codes suggests that approximately 34 (36-2) codes remain to be elicited. The estimate of total number of codes is intended for use in arranging and budgeting for the remaining number of interviews. The estimate is not intended to be the definitive statistical estimate of number of codes.

One key assumption for both Kaplan-Meier and capture-recapture is independence. This is unlikely to be true with items (codes) for a patient reported outcome where items are routinely assessed for their correlation. Preliminary assessments, using capture-recapture methods such as the Lincoln Peterson, or Chapman estimator were prepared but not presented here. The estimators were computed at each interview. Lincoln-Peterson and Chapman tended to underestimate the true number of codes. The Good-Turing (GT) is simple to compute and underestimated the true number for the initial 5- interviews, then overestimated. For example, while the true known number of codes is 19, after 5 interviews the GT estimator was 19.5 and after 6 interviews was 28.4. Hypothetically, If the estimator at 5 interviews was correct, then approximately 14 (19-5) or fewer interviews remain. Again the discrepancies are likely due to the induced dependence among codes elicited at interviews

## References

Food and Drug Administration, 2009 Guidance for Industry, Patient-Reported Outcome Measures: Use in Medical Product Development to Support Labeling Claims, https://www.fda.gov/media/77832/download, accessed 2020.

Bernard, Russel, editor field methods, email of Feb 8, 2020.

Brittain, Sarah, and Dankmar Böhning. "Estimators in capture–recapture studies with two sources." *AStA Advances in Statistical Analysis* 93.1 (2009): 23-47.

Bowen GA. Naturalistic inquiry and the saturation concept: a research note. Qualitative Research 2008;8(1):137-52.

Brédart, Anne, et al. "Interviewing to develop Patient-Reported Outcome (PRO) measures for clinical research: eliciting patients' experience." *Health and quality of life outcomes* 12.1 (2014): 15.

Chun, Y.H., 2006. Estimating the number of undetected software errors via the correlated capture–recapture model. *European journal of operational research*, *175*(2), pp.1180-1192.

Francis, Jill J., et al. "What is an adequate sample size? Operationalising data saturation for theory-based interview studies." *Psychology and Health* 25.10 (2010): 1229-1245.

Fusch, Patricia I., and Lawrence R. Ness. "Are we there yet? Data saturation in qualitative research." The Qualitative Report 20.9 (2015): 1408.

Guest, Greg, Arwen Bunce, and Laura Johnson. "How many interviews are enough? An experiment with data saturation and variability." *Field methods* 18.1 (2006): 59-82.

Good, Irving J. "The population frequencies of species and the estimation of population parameters." *Biometrika* 40.3-4 (1953): 237-264.

Kaplan, E. L., & Meier, P. (1958). Nonparametric estimation from incomplete observations. Journal of the American statistical association, 53(282), 457-481.

Kleinman, Leah, et al. "The anemia impact measure (AIM): development and content validation of a patient-reported outcome measure of anemia symptoms and symptom impacts in cancer patients receiving chemotherapy." Quality of Life Research 21.7 (2012): 1255-1266.

Mason, Mark. "Sample size and saturation in PhD studies using qualitative interviews." Forum qualitative Sozialforschung/Forum: qualitative social research. Vol. 11. No. 3. 2010.

Orians, Gordon H., and P. H. Leslie. "A capture-recapture analysis of a shearwater population: with a statistical appendix." The Journal of Animal Ecology (1958): 71-86.

O'Reilly M, Parker N. 'Unsatisfactory Saturation': a critical exploration of the notion of saturated sample sizes in qualitative research. Qualitative Research 2013;13(2):190-7.

Patrick, Donald L., et al. "Content validity—establishing and reporting the evidence in newly developed patient-reported outcomes (PRO) instruments for medical product evaluation: ISPOR PRO Good

Research Practices Task Force report: part 2—assessing respondent understanding." Value in Health14.8 (2011): 978-988.

Patrick, Donald L., et al. "Content validity—establishing and reporting the evidence in newly developed patient-reported outcomes (PRO) instruments for medical product evaluation: ISPOR PRO good research practices task force report: part 1—eliciting concepts for a new PRO instrument." Value in Health 14.8 (2011): 967-977.

R Core Team (2020). R: A language and environment for statistical computing. R Foundation for Statistical Computing, Vienna, Austria.
URL https://www.R-project.org/.

Seale C. Grounding theory. In: Seale C, editor. The Quality of Qualitative Research.London: SAGE Publications Ltd; 1999. p. 87-105.

Stolper , http://www.gutfeelings.eu/glossary/saturation-2/   accessed 2017.

Kate Tilling, Capture-recapture methods—useful or misleading?, *International Journal of Epidemiology*, Volume 30, Issue 1, February 2001, Pages 12–14,

van Rijnsoever FJ (2017) (I Can't Get No) Saturation: A simulation and guidelines for sample sizes in qualitative research. PLoS ONE12(7): e0181689. https://doi.org/10.1371/journal.pone.0181689

Wittes, J.T., Colton, T. and Sidel, V.W., 1974. Capture-recapture methods for assessing the completeness of case ascertainment when using multiple information sources. *Journal of chronic diseases*, *27*(1), pp.25-36.

Wood, Fred, and Tom Guinter. "Evolution and implementation of the CDISC study data tabulation model (SDTM)." *Pharmaceutical Programming* 1.1 (2008): 20-27.

Wu, Jianrong. "Single-arm phase II cancer survival trial designs." Journal of biopharmeutical statistics 26.4 (2016): 644-656.